\documentclass[prc,aps,twocolumn,floats,floatfix,showpacs,superscriptaddress,nofootinbib]{revtex4}
\usepackage{amsmath}
\usepackage{amsfonts}
\usepackage{graphicx}

\begin{document}

\title{
Spreading widths of   giant resonances in
spherical nuclei:

damped transient response
}

\author{A. P. Severyukhin}
\affiliation{Bogoliubov Laboratory of Theoretical Physics,
	Joint Institute for Nuclear Research,
	141980 Dubna, Moscow region, Russia}
\affiliation{Dubna State University, 141982 Dubna, Moscow region, Russia}
\author{S. {\AA}berg}
\affiliation{Mathematical Physics, Lund University,
	PO Box 118, S-22100 Lund, Sweden}
\author{N. N. Arsenyev}
\affiliation{Bogoliubov Laboratory of Theoretical Physics,
	Joint Institute for Nuclear Research,
	141980 Dubna, Moscow region, Russia}
\author{R. G. Nazmitdinov}
\affiliation{Departament de F\'{\i}sica, Universitat de les Illes Balears,
	E-07122 Palma de Mallorca, Spain}
\affiliation{Bogoliubov Laboratory of Theoretical Physics,
	Joint Institute for Nuclear Research,
	141980 Dubna, Moscow region, Russia}
\affiliation{Dubna State University, 141982 Dubna, Moscow region, Russia}
\begin{abstract}
We propose the universal approach to describe spreading widths
of monopole, dipole and quadrupole  giant resonances in heavy and superheavy
spherical nuclei. Our approach is based on the ideas of the random matrix distribution of the
coupling between one-phonon and two-phonon states generated in the
random phase approximation.
We use the Skyrme interaction SLy4 as our model Hamiltonian to create a
single-particle spectrum and to analyze excited states of the doubly
magic nuclei $^{132}$Sn, $^{208}$Pb and $^{310}$126.
Our results demonstrate that the universal approach enables to describe
gross structure of the spreading widths of the considered  giant resonances.
\end{abstract}


\date{\today}

\maketitle
%
%
Damping of collective motion in finite many-body quantum systems is
among topical subjects in mesoscopic physics.
The question of how, for example,
multipole giant resonances ~(GRs) in
nuclei ~\cite{dstadta} and metal clusters ~\cite{ReiS}
dissolve their energy is still not well understood.
There is, however, a consensus of opinion that,
in particular, in a nucleus, once excited by an external field, a
GR progresses to a fully equilibrated system via
direct particle emission and by coupling to more complicated
states produced by the intrinsic motion of nucleons
(see, for example, Ref.\cite{bert}).
The former mechanism gives rise to escape width $\Gamma_p$. It is expected that the decay
evolution along the hierarchy of more complex configurations till
compound states determines spreading width $\Gamma$.
A full description of
this decay represents a fundamental problem which is, however,
difficult to solve (if even  is possible at all ?) due to
existence of many degrees of freedom.

In general, the description of spreading width in mesoscopic systems is based on the
study of the electromagnetic strength distribution (strength function)~\cite{BM}
in some energy interval. This interval
 should be large enough to catch hold of basic features of a GR
under investigation. Note, that in deformed systems the experimental widths
are systematically larger and may develop a two- or three-peak structure.
In this paper we consider only spherical nuclei
in order to highlight  a generic nature of
the width $\Gamma$ in monopole, dipole and quadrupole resonances
in heavy and super-heavy systems.

Nuclear shell model may be used to analyse spreading widths of GRs.
However, the complexity of the calculations increases rapidly with the size of the
configuration space. This fact severely restricts the feasibility of
shell model calculations for heavy and super-heavy nuclei.
In addition, even for a medium $^{48}$Ca isotope the state-of-art shell model
calculations \cite{jap}, which operate with the Hamiltonian matrices of
a huge dimension, produce questionable results for the dipole GR.
Although these calculations reproduce reasonably well its peak position and peak
width, the enhancement of the classical Thomas-Reiche-Kuhn sum rules is
too overestimated. As a result, the number of
shell model studies, in particular, dipole GRs in heavy and super-heavy
nuclei are limited and rather focused on details of
low-energy region (e.g., \cite{dres}).

The success of random matrix theory (RMT)~\cite{B81,M91,zel,GMW,HW,Gom},
based on universal features in spectra of complex quantum systems,
gives hope to shed light on the spectral properties
and the distribution of transition-strength properties of GRs, when
specific details become not of a primary importance.
As is well known,
the RMT assumes only that a many-body Hamiltonian belongs to an ensemble
of random matrices that are consistent with the fundamental
symmetries of the system such as parity, rotational, translational and
time-reversal symmetries.
We belive that it is quite suitable
for our aim: to provide a generic principle for the decay  of
highly excited states with
angular momentum and parity: $J^\pi=0^+,1^-,2^+$.
On the other hand, to understand the realistic
fragmentation of high-lying states over complex configurations,
observed as the spreading width, it is necessary also to exploit
a realistic nuclear structure model. It should be based on the microscopic
many-body theory, where the effects of the residual interaction on
the statistics must be studied in large model spaces. Introducing
a residual interaction in general implies a transition to
 the Gaussian orthogonal ensemble (GOE)
-properties above some excitation energy~\cite{A}.
In fact, recent analysis of 151 experimental nuclear levels up to excitation energy of
$E_x=6.2$ MeV in $^{208}$Pb indicates already that the spectral properties
are described by the GOE due to a residual interaction, even though there is
a small admixture of regular dynamics brought about by
the low-lyings states \cite{mun}.

The quasiparticle-phonon model (QPM)~\cite{S} offers an attractive
framework for such studies.  We will use the
modern development of the QPM, a finite rank separable
approximation (FRSA)~\cite{gsv98}. That approach
employs the Skyrme forces
to calculate  the single-particle~(sp) spectrum
and the residual interaction in a self-consistent manner
in order to avoid any  artefacts~\cite{BNY}.
As an example of the parameter
set, we consider widely used SLy4~\cite{sly} which is adjusted to
reproduce the nuclear matter properties, as well as nuclear charge
radii, binding energies of doubly magic nuclei.
This set shifts the island of stability
towards high charge
numbers around $^{310}_{184}$126~\cite{b00}.
Evidently, another parameter
set can be used as well for our purposes.
The continuous part of the sp  spectrum is discretized by diagonalizing
the Hartree-Fock Hamiltonian on a harmonic oscillator basis. The cut-off of
the continuous part is at the energy of 100~MeV.

The residual particle-hole interaction is obtained as the second derivative of
the energy density functional with respect to the particle
density. By means of the standard procedure~\cite{tebdns05} we
obtain the familiar equations of the random phase approximation (RPA)
in the one particle-one hole (\rm{1p-1h}) configuration
space. The eigenvalues of the RPA equations are found numerically
as the roots of a relatively simple secular equation within the
FRSA~\cite{gsv98}. Being a linear combination of 1p-1h states,
the RPA solutions are treated as quasi-bosons with quantum numbers $\lambda^\pi$.
Among these solutions there are one-phonon states corresponding to collective GRs and pure
two-quasiparticle states.
The configurations with various degree of
complexity can be built by combining different one-phonon configurations
$\lambda_1^{\pi_1},\lambda_2^{\pi_2},\cdots$ of fixed quantum number $\lambda^\pi$.
As a result, one obtains the n-phonon components
$[\lambda_1^{\pi_1}\otimes\lambda_2^{\pi_2}\otimes\cdots\lambda_n^{\pi_n}]_{\lambda^\pi}$
of the wave function.
The diagonalization of the model Hamiltonian in the space of  the one-phonon
and complex configurations produces eigenstates of excited states. These states
carry information on the fragmentation of one-phonon component over
complex configurations in the resulting
eigenfunction.

A natural question arises: what degree complexity of configuration
should be enough in order to understand a gross
structure of a particular GR which data are available  in modern
experiments? In addition, once this complex configuration is
defined one can further ask about statistical properties of states
that compose a GR strength distribution.

In the actual calculations of the GRs strength distributions
in spherical nuclei  $^{132}$Sn, $^{208}$Pb and $^{310}$126
considered as examples, we have included in our
model space different multipoles $\lambda^\pi=0^+,1^-,2^+,3^-,4^+$.
Tentative estimates for the position of the resonance centroids
$E_c$ and the spreading width $\Gamma$ have been defined
by means of the energy-weighted moments $m_k=\sum_{}B(E\lambda)\,{E}^k$:
i) $E_{\rm c}=m_1/m_0$;
ii) $\Gamma=2.35\sqrt{m_2/{m_0}-\left(m_1/{m_0}\right)^2}$
(see, for example, \cite{lip}).
Note that the coefficient $2.35$ has its roots to the experimental
definition of the  width (full width at half maximum) related to the
variance of the Gaussian (see, for example, \cite{bert}).
Next, we construct  various
combinations $\omega_{\lambda_1i_1}+\omega_{\lambda_2i_2}$
to define the energy interval for location of the resonance width
of fixed quantum number $\lambda^\pi$,
taking 95{\%} of the energy-weighted sum rule symmetrically
around the centroid's position ($E_{\rm c}$).
It is noteworthy that for all GRs, considered in the present paper,
the matrix elements for direct excitation of two-phonon components from the ground state
are about two orders of magnitude smaller relative to ones for the
excitation of one-phonon configurations.
On the other hand, the density of these complex configurations
is much higher than the one-phonon ones and contributes essentially to
statistics of the final states.

From our preliminary analysis of  complex structure observed
in the region of the isoscalar giant monopole resonance~(ISGMR)
with $J^\pi=0^{+}$ of the doubly magic nucleus $^{208}$Pb \cite{p1}
we have found that the spectrum can be explained as a result
of mixing of one- and two-phonon components of the wave function, i.e.,
\begin{eqnarray}
\label{wf}
&&\Psi_\nu(JM)=\biggl\{\sum\limits_i R_i(J\nu)Q_{JMi}^{+}+\\
&&\sum\limits_{\lambda_1i_1\lambda_2i_2}P_{\lambda_2i_2}^{\lambda_1i_1}(J\nu)
\left[Q_{\lambda_1\mu_1 i_1}^{+}Q_{\lambda_2\mu_2
i_2}^{+}\right]_{JM}\biggr\}\mid 0\rangle\,,\nonumber
\end{eqnarray}
where $Q_{\lambda\mu i}^{+}|0\rangle$ is the RPA excitation having
energy $\omega_{\lambda i}$; $\lambda$ denotes the total angular
momentum and $\mu$ is its z-projection in the laboratory system.

In the case of the phonon-phonon coupling (PPC)
the variational principle leads to a set of linear equations for
the unknown amplitudes $R_i(J\nu)$ and
$P_{\lambda_2i_2}^{\lambda_1i_1}(J\nu)$~
(see details in Ref.\cite{svg04}):
\begin{equation}
 \label{2pheq1}
(\omega _{Ji}-E_\nu )R_i(J\nu )+\sum_{\lambda _1i_1 \lambda_2i_2}
U_{\lambda _2i_2}^{\lambda _1i_1}(Ji) P_{\lambda_2i_2}^{\lambda
_1i_1}(J\nu )=0,
\end{equation}
\begin{equation}
 \label{2pheq2}
 \sum\limits_iU_{\lambda _2i_2}^{\lambda _1i_1}(Ji)R_i(J\nu ) +
2(\omega _{\lambda _1i_1}+\omega _{\lambda _2i_2}-E_\nu )
P_{\lambda _2i_2}^{\lambda _1i_1}(J\nu)=0.
\end{equation}
To resolve this set it is required to compute the coupling matrix elements
\begin{equation}
\label{cme}
U_{\lambda _2i_2}^{\lambda _1i_1}(J i) = \langle 0| Q_{J i } H
\left[ Q_{\lambda _1i_1}^{+}Q_{\lambda _2i_2}^{+}\right] _{J} |0\rangle
\end{equation}
between one- and two-phonon configurations.
Our approach is similar to the
particle-vibration coupling (PVC) model based on Green's function method
(see for a recent  review Ref.\cite{bal}) that has been used in the study of the monopole
\cite{bor1} and the quadrupole \cite{bor2}
GR
widths in $^{208}$Pb with the aid of Skyrme forces.
Note, that the PPC
includes as well the coupling of one-phonon state with
two particle-two hole states, important in the PVC model, as a particular case
(see discussion in Chapter 4.3 of the textbook \cite{S}).

\begin{figure}[t!]
	\includegraphics[width=0.7\columnwidth]{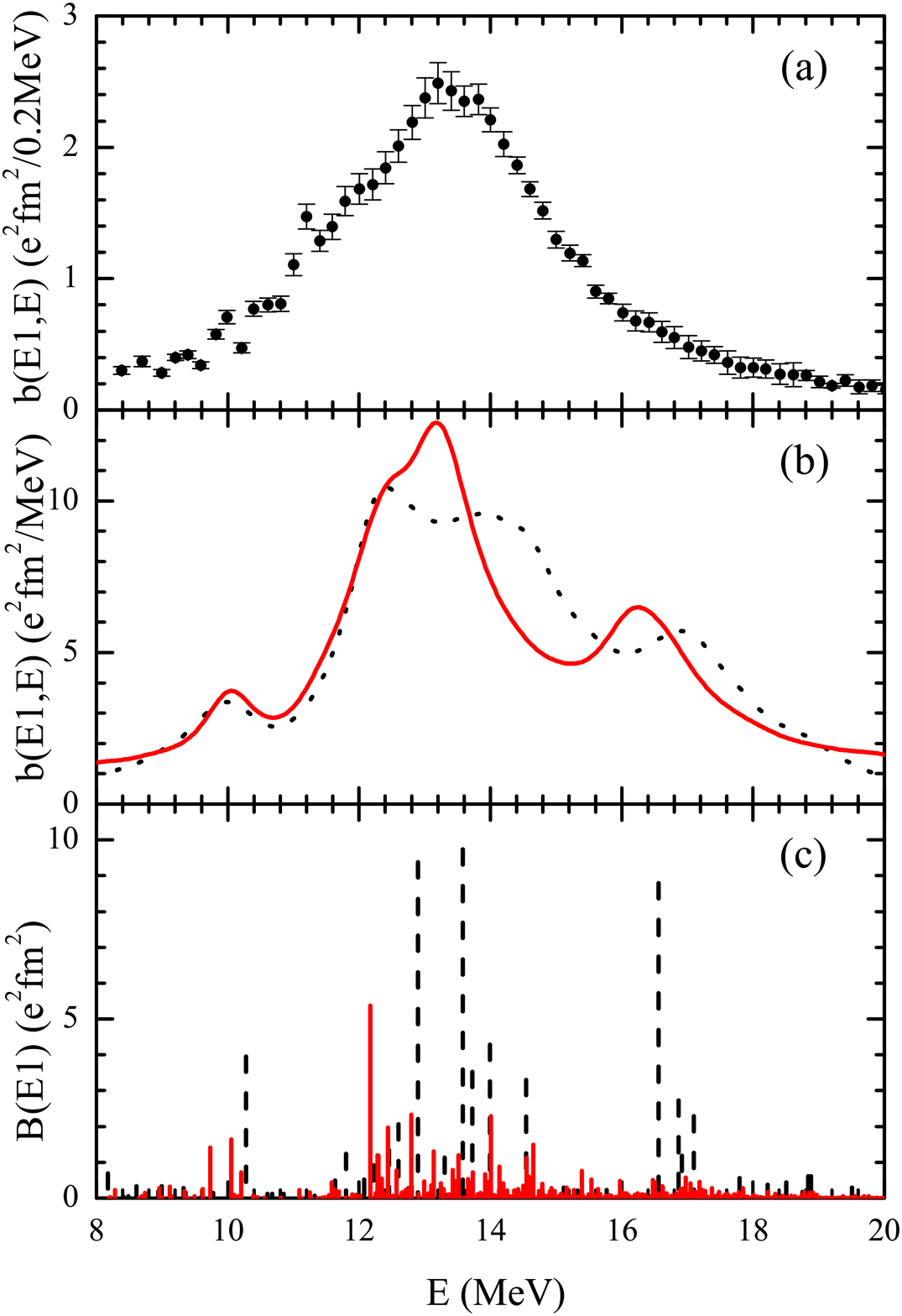}
	\caption{(Color online) $^{208}$Pb: (a) experimental $B(E1)$ strength distribution;
		(b) the comparison of the results obtained by means of the microscopic
        (dotted line) and
		the random (solid line) coupling matrix elements between the one- and
		two-phonon configurations;
		(c) $B(E1)$ strength distribution for one-phonon states (dashed
		line) and for the PPC case.
		The smoothing parameter 1 MeV is used for the strength distribution described
		by  the Lorentzian function.
		 The experimental
		data are taken from Ref.~\cite{TP11}}\label{fig1}
\end{figure}

We start our discussion from the analysis of the spreading width of
the Isovector Giant Dipole Resonance (IVDGR) in the spherical $^{208}$Pb nucleus,
since it is the best known example of
nuclear vibrations. The coupling (the PPC) of the one-phonon states with an intermediate
complex background of two-phonon states yields a strong redistribution of the
 one-phonon dipole strength in the region of the
IVDGR (see Fig.1c). It suppresses the high-lying one-phonon strength near $(\sim 17$ MeV) and
pushes this strength down (see also \cite{g}).
As a result, we obtain a reasonably well description of the dipole strength
distribution over the resonance localization region (compare Figs.1a,b).
It appears that the presence of two-phonon components
in our wave function, in addition to the one-phonon ones, already
enables us to describe the gross strength distribution of the
typical dipole response in the heavy spherical nucleus $^{208}$Pb.
Similar conclusions have been drawn on the basis of shell-model calculations
for the states above 8 MeV in Ref.\cite{dres}.

The relatively broad realistic distribution seen in Fig.\ \ref{fig1} indicates
that many two-phonon configurations contribute to the fragmentation process.
Indeed, the RMT measures such as the nearest-neighbor spacing distribution (NNSD) and the spectral rigidity $\Delta_{3}$ indicate a transition towards the GOE when the coupling is switched on
(see Figs.3,4 in \cite{p1} for the ISGMR). Evidently, the extension of the wavefunction to more
complex configurations would increase the fragmentation of the one-phonon strength
over many excited states. This complexity suggests an approach from random matrix theory to
describe the fragmentation of the transition strength between the RPA states and the ground state.

The coupling of the phonon states to more complex background states can be described by a simple doorway
state Hamiltonian (cf Ref.\ \cite{BM})
\begin{equation}
H_{J^\pi}=H_d+H_b+V
\end{equation}
where $H_d$ describes the doorway states, $H_b$ the background states and $V$ the coupling between doorway
states and background states.
The RPA-phonon states constitute the doorway states,  $H_d=\sum_i \omega_{Ji} Q^+_{Ji}Q_{Ji}$,
and the background states are two-phonon and possibly more complex states, with eigenstates,
$H_d|d\rangle=\omega_d |d\rangle$ and $H_b|b\rangle=e_b|b\rangle$, respectively.
The Hamiltonian $H_{J^\pi}$ represents a set of good
quantum numbers, $J^{\pi}$, and the RPA phonons as well as all background states
fulfill these quantum numbers. We assume
no coupling between different doorway states or between different background states,
$\langle d|V|d' \rangle=0$ and $\langle b|V|b'\rangle =0$, but all coupling takes place between
the doorway states and the complex background states, $\langle d |V| b \rangle=V_{db}$.
Similar ideas have been discussed in \cite{zel2} where some limiting analytical
estimates  were obtained for the GDR strength function.

In our consideration the doorway states are taken from
the microscopic RPA calculation for the isoscalar or isovector
$J^\pi$ mode providing energies, $\omega_d$, and transition matrix elements to the ground state ($|0\rangle$), $B_d=\langle d|\mathcal{M}_{J^{\pi}}| 0 \rangle$, via the transition operator  $\mathcal{M}_{J^{\pi}}$.
No transition can occur between a background state and the ground state, $0=\langle b|\mathcal{M}_J^{\pi})|0 \rangle$.
After diagonalisation of the Hamiltonian $H_{J^\pi}$,
the transition strength of the doorway states is fragmented
on all states, and provides the full transition strength distribution.

The modelling of background states, $e_b$, and couplings, $V_{db}$, can be performed on
different levels of approximation. In the PPC model, the background energies $e_b$ are obtained as the sum of two RPA phonon energies, $\omega_{\lambda_1}+\omega_{\lambda_2}$,  coupled to $J^\pi$.
The coupling matrix elements, $V_{db}$, are subsequently obtained from Eq.(\ref{cme}).
To account for underlying complexity we now replace these matrix elements by a random coupling.
The parameter that determines the strength of the coupling is the rms
value of the matrix elements, $\sigma=\sqrt{\langle V_{db}^2\rangle}$.
The actual distribution of the random interaction is not important, as long as it is symmetric, $\langle V_{db}\rangle=0$. While the microscopic matrix elements follow a truncated Cauchy distribution,
we chose a Gaussian distribution for the random interaction,
\begin{equation}
\label{rand}
P(V)=\frac{1}{\sigma\sqrt{2\pi}} \exp{\frac{-V^2}{2\sigma^2}}.
\end{equation}
Solutions of $H_{J^{\pi}}$ are ensemble averaged over the random interaction and give the transition strength distribution.

Choosing the strength of the random interaction from the microscopic calculation of coupling matrix elements, that is the rms value, {\bf $\sigma_c,$} of the coupling matrix elements given by
Eq.(\ref{cme}), we
get the $B(E1)$ distribution strength of IVGDR for $^{208}$Pb as shown in Fig.\ \ref{fig1}b.
It is noteworthy that the comparison of the strength distributions obtained
with the aid of the PPC and the random distribution of the matrix elements
demonstrates a remarkable similarity (see Fig.1b).
Moreover, by means of the latter distribution (\ref{rand}) we reproduce
the experimental strength distribution of the IVGDR as well (compare Figs.1a,b).

\begin{figure}[t!]
	\includegraphics[width=0.8\columnwidth]{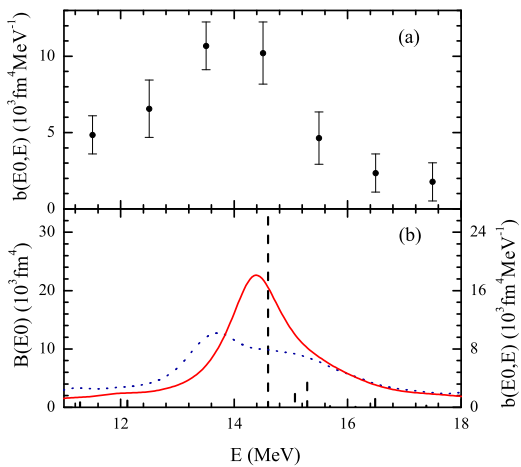}
	\caption{(Color online) The same as Fig.~\ref{fig1}, in the case
		of ISGMR in $^{208}$Pb.
        The experimental data are taken from
		Ref.~\cite{P13}}\label{fig2}
\end{figure}

The RPA analysis provides the location of the ISGMR in  $^{208}$Pb
in the energy region $E_{x}=10.5-18.5$~MeV.
The PPC yields a detectable redistribution of the ISGMR strength in
comparison with the RPA results. It results in
the 1~MeV downward shift of the main peak (see Fig.\ \ref{fig2}).
Our analysis shows that the major
contribution to the strength distribution is brought about by the
coupling between the $[0^+]_{RPA}$ and $[3^-\otimes3^-]_{RPA}$
components.
In contrast, the use of the random matrix distribution yields the
backshifting of this peak. Evidently, in this case there is
only an average strength that does not produce any preferences in
the coupling between one- and two-phonon states
of different one-phonon nature.
The strength distribution of the ISGMR
obtained in this case is rather close to the experimental
distribution~\cite{P13}, see Fig.\ \ref{fig2}.

In the same manner we calculate and compare different estimations for
the strength distribution of the GRs  in $^{132}$Sn and $^{310}$126 nuclei.
The results of calculation and comparison with the experimental data
and the empirical systematics are displayed in Table~\ref{tab1}.
The description of the spreading width by means of the PPC and
the random distribution (\ref{rand}) provide similar results for the
ISGMR and IVGDR in all considered nuclei. For isoscalar giant quadrupole
resonance (ISQGR) the PPC yields the widths that are larger relative
to the ones produced by the random distribution.
It is required reliable experimental
measurements in order to remove systematic uncertainties in experimental
analysis based on optical potentials (see also the discussion in Ref.\cite{bor2}).

Considering the interaction strength as a parameter, we investigate the complexity of the
energy states in terms of the NNSD by studying the Brody mixing parameter \cite{B81}, $q$, versus $\sigma$. A smooth increase is found
from regularity ($q = 0$;  Poisson statistics) when $\sigma=0$ to
chaos ($q \approx 1$; GOE) when $\sigma=\sigma_r$,
where the critical value, $\sigma_r$, depends on considered nucleus and
kind of GR.
 It is remarkable that the onset of chaos appears at a $\sigma$-value
very similar to the interaction strength of the microscopic phonon coupling model.
We thus find that $\sigma_c \approx \sigma_r$ for each considered case.
A way to chose the strength of the random interaction may thus be
to find the $\sigma$ value where the GOE properties appear, $\sigma_r$,
(practically defined as $q$=0.95) rather than performing the full microscopic PPC calculation.

While the NNSD provides information about correlations
on short energy scales, the spectral-rigidity meassure $\Delta_3$
characterises long-range correlations between the energy levels.
For the coupling strength\ $\sigma_c$\ $(\approx \sigma_r)$ full
short-range GOE correlations were found in the NNSD.
The spectral rigidity $\Delta_{3}$ only reproduces the
GOE distribution $\bar{\Delta}_{3}(L)\approx \frac{1}{\pi^2}(\ln L -0.0687)$
up to a $L$-value, $L_{\text{max}}$. For the IVGDR of $^{208}$Pb we find
$L_{\text{max}}$=7. This implies long-range
GOE correlations in the strength distribution around the centroid
energy within an energy range of about $L_{\text{max}}/\rho(E_c)=0.2$ MeV, where $\rho$ is the
density of background states.
Consequently, only correlations beyond this energy range may provide specific structure information.
Note, however, that the smoothing (1 MeV) smears out the strength effectively over more background states,
not considered in the model. As a result, the correlation energy obtained by means of
$L_{\text{max}}$ expected to  be larger.

Since the energy spectrum shows  full GOE properties when the appropriate coupling
strength has been included, another step in the doorway state model can be
introduced. Instead of calculating the background state energies
with the aid of the RPA calculations,
one might employ
random GOE-generated energies,
following a smooth level density function of background states. The resulting strength distribution
calculated in this way coincides perfectly with the case when microscopic background energies are included.
This further simplifies the model, and provides possibilities to calculate spreading widths
of giant resonances in a quite universal way.

 \begin{table*}
 \caption{Characteristics of the Giant Multipole Resonances for
 $^{132}$Sn, $^{208}$Pb and $^{310}$126 nuclei:
 centroid energies $E_c$ and the spreading widths $\Gamma$
 calculated with the RPA and RPA plus phonon-phonon coupling with
 the microscopic (PPC) and random distribution of coupling matrix elements
 (Random), are compared with available experimental
 data~\cite{P13,AK05,BF75,YL04}. The values of $E_c$ and
 $\Gamma$ have been computed in corresponding energy intervals
$\Delta{E}$. For comparison the centroid energy and width values
from the empirical systematics (Syst.) are presented~\cite{Au75,BF75,BE81}.}
\label{tab1}
\begin{ruledtabular}
\begin{tabular}{ccccccccccccccc}
&  &\multicolumn{5}{c}{$E_c$ (MeV)}         &\multicolumn{5}{c}{$\Gamma$ (MeV)}  &$\Delta{E}$ (MeV)\\
&  &Expt.&Syst.&\multicolumn{3}{c}{Theory}&Expt.&Syst.&\multicolumn{3}{c}{Theory}&\\
&  &     &     &RPA&PPC&Random            &     &     &RPA&PPC&Random            &\\
\noalign{\smallskip}\hline\noalign{\smallskip} ISGMR
&$^{132}$Sn& --           &15.71&16.8&16.6&16.7  &             & -- &2.7&4.7&3.4 &12-21\\
&$^{208}$Pb&$13.7{\pm}0.1$&13.50&14.7&14.4&14.6  &$3.3{\pm}0.2$& -- &1.9&3.9&2.7 &10.5-18.5\\
&          &$13.96{\pm}0.20$&   &    &    &      &$2.88{\pm}0.20$&  &   &   &    & \\
&$^{310}$126& --          &11.82&12.7&12.7&12.7  &             & -- &1.4&3.0&1.8 &9.5-16\\
\noalign{\smallskip}\hline\noalign{\smallskip} IVGDR
&$^{132}$Sn&$16.1{\pm}0.7$&15.26&15.5&15.4&15.3  &$4.7{\pm}2.1$&4.67&4.9&5.0&5.2 &11-20\\                         &$^{208}$Pb&13.43         &13.73&14.0&14.0&13.8  & 4.07        &4.15&4.6&4.9&4.8 &9.5-18.5\\                      &$^{310}$126& --          &12.53&12.6&12.6&12.4  & --          &3.80&4.3&4.4&4.7 &8.5-17.5\\
\noalign{\smallskip}\hline\noalign{\smallskip} ISGQR
&$^{132}$Sn& --           &12.71&14.8&14.7&14.7  & --          &3.91&2.1&4.0&2.6 &10-20\\                         &$^{208}$Pb&$10.89{\pm}0.30$&10.92&13.0&13.0&13.0&$3.0{\pm}0.3$&3.04&1.4&3.1&2.1 &8-18\\                          &$^{310}$126& --          &9.56&11.5&  11.4  &  11.4     & --          &2.43&1.2& 2.7  &  1.7  &8-16\\                          \end{tabular}                                                                                                     \end{ruledtabular}                                                                                                \end{table*}

In summary, we suggest the way to describe spreading widths of GRs
by including the coupling between one-phonon and two-phonon states.
This coupling can be generated by means
of the random distribution of coupling matrix elements
(\ref{rand}), and the energies of the two-phonon states can be generated from
the GOE distribution. The variance of the Gaussian function $\sigma^2$
can be obtained from the GOE limit of the NNSD
of spectra generated by the coupling between one- and
two-phonon states, characterised by the same quantum number $J^\pi$.
\\

%
\section*{Acknowledgments}
The authors thank Nguyen Van Giai, V. Yu. Ponomarev and H. Sagawa
for useful discussions.
A.P.S. thanks for the hospitality at the Division of Mathematical
Physics, Lund University, where a part of this work has been done.
S. \AA. thanks the Swedish Natural Science Research Council for financial
support, and the Bogoliubov Laboratory of Theoretical Physics for warm
hospitality.
This work was partly supported by Russian Foundation for Basic
Research under Grant nos. 16-52-150003 and 16-02-00228.

%
%

\end{document}